\documentclass[a4paper, 12pt, notoc]{JHEP}
\usepackage{latexsym}
\def\cA{\mathcal{A}}
\def\cB{\mathcal{B}}

\def\cF{\mathcal{F}}

\def\cT{\mathcal{T}}

\def\mC{{\bf C}}
\def\mP{{\bf P}}
\def\mZ{{\bf Z}}
\def\chern{\textrm{ch}}
\def\ch{\textrm{c}}
\def\todd{\textrm{td}}
\def\rank{\textrm{rk}}

\def\tr{\textrm{tr}}
\def\mod{\textrm{ mod }}
\def\bb1{\textup{\small{1}} \kern-3.8pt \textup{1}}
\newcommand{\diff}[2]{\textrm{d}^{#1}{#2}}
\newcommand{\BS}[1]{|\; #1 \;\rangle\rangle_B}

\title{D-branes on some one- and two-parameter Calabi-Yau hypersurfaces}
\author{Emanuel Scheidegger\thanks{Resarch supported in part by Deutsche
Forschungsgemeinschaft and by the European Commission TMR programme
ERBFMRX-CT96-0049}\\
\textit{Sektion theoretische Physik der}\\
\textit{Ludwig-Maximilian-Universit\"at M\"unchen}\\
\textit{80333 M\"unchen, Germany}\\
\email{esche@theorie.physik.uni-muenchen.de}}
\abstract{
  D-branes on one-parameter Calabi-Yau spaces and two-parameter
$K3$-fibered Calabi-Yau 
manifolds are analyzed from both the Gepner model point of view and the
geometric perspective. 
We compute part of the spectrum of the boundary states and comment on
the appearance of the 
$D0$-brane as well as on nonsupersymmetric large volume configurations
becoming supersymmetric
 at the Gepner point.
}
\preprint{hep-th/9912188}
\begin{document}
\section{Introduction}
D-branes (for review, see eg.~\cite{9611050}) play a pivotal r\^ole 
for string dualities. Most work to date has been on D-branes
in flat space-time. However, a thorough understanding of string
dualities
in four dimensions, requires D-branes wrapped on cycles of the
compactifying 
manifold. Of those, Calabi-Yau manifolds are of particular interest. 
The problem of D-branes wrapped on cycles of compact Calabi-Yau three
folds 
was the subject of two very interesting papers~\cite{9906200,9910172}.
In the 
first of these papers the 
simplest Calabi-Yau three-fold, the quintic hypersurface in 
four-dimensional projective space has been studied in detail. 
In particluar the question whether the geometric picture of 
D-branes wrapping on cycles survives in the 
stringy regime was addressed. In the second paper Diaconescu and 
R\"omelsberger have extended relevant aspects of this analysis
to an elliptically fibred two-parameter Calabi-Yau manifold. 
The same question has been studied also in~\cite{9907131} and for a
non-compact 
Calabi-Yau manifold in~\cite{9906242}. 
The general picture has been expounded by Douglas~\cite{9910170}. 
Even though in these papers important progress has been made, 
D-branes on Calabi-Yau spaces are far from being understood. 
The aim of this paper is to apply some aspects of the analysis
of~\cite{9906200,9910172} 
to a few more models. 
These are the three remaining one-parameter 
models and the two two-parameter K3 fibrations in weighted projective
space. \\

In section~\ref{sec:one} we provide some relevant background material
about the 
one-parameter models and compute the intersection matrix on the
three-cycles on the 
mirror as a preparation for the comparison with the computations from
superconformal 
field theory. This will then be done for the $K3$ fibrations in
section~\ref{sec:two}. We then 
provide in section~\ref{sec:periods} the basic tool in relating the
spectrum of supersymmetric brane 
configurations at the large volume limit to the periods of the
holomorphic three-form 
of the Calabi-Yau manifold. In particular we compute the number of
moduli of branes 
wrapped on the $K3$ fiber. In section~\ref{sec:Gepner} we discuss the
boundary conformal field theory 
for the Gepner models associated with the Calabi-Yau manifolds under
consideration. 
Finally in section~\ref{sec:D-branes} we put everything together and
discuss some interesting boundary states. \\  

{\bf Note:} The part on the two-parameter $K3$-fibrations has
considerable overlap with 
the recent preprint~\cite{9912147} which appeared while we were in the
process of 
finalizing this paper. Where applicable, our results agree.

\section{The 1-parameter models}
\label{sec:one}

There are four 1-parameter models of Fermat type: First, there is
$\mP_{1,1,1,1,1}^4[5]$ 
known as the quintic in $\mP^4$ which has been extensively studied
in~\cite{NPB359-21} 
and~\cite{9906200}. Then there are the degree 6 hypersurfaces
$\mP_{1,1,1,1,2}^4[6]$ in 
the weighted projective space $\mP_{1,1,1,1,2}^4$, the degree 8
hypersurfaces 
$\mP_{1,1,1,1,4}^4[8]$ in $\mP_{1,1,1,1,4}^4$ and the degree 10
hypersurfaces 
$\mP_{1,1,1,1,4}^4[8]$ in $\mP_{1,1,1,2,5}^4$ which were explored from
the point 
of view of mirror symmetry in~\cite{9203084} and~\cite{9205041}. Recall
that the 
weighted projective space $\mP_{w_1,w_2,w_3,w_4,w_5}^4$ is defined by
\begin{equation}
\mP_{w_1,w_2,w_3,w_4,w_5}^4 = \frac{\mC^5\setminus
\{0\}}{(z_1,z_2,z_3,z_4,z_5) 
\sim
(\lambda^{w_1}z_1,\lambda^{w_2}z_2,\lambda^{w_3}z_3,\lambda^{w_4}z_4,\lambda^{w_5}z_5)}
\end{equation}
The hypersurfaces $X_i$ are typically given by
\begin{eqnarray}
X_1:&z_1^6 + z_2^6 +z_3^6 +z_4^6 +z_5^3 = 0,&(z_1:z_2:z_3:z_4:z_5) \in 
\mP_{1,1,1,1,2}^4\\
X_2:&z_1^8 + z_2^8 +z_3^8 +z_4^8 +z_5^2 = 0,& (z_1:z_2:z_3:z_4:z_5) \in
\mP_{1,1,1,1,4}^4\\
X_3:&z_1^{10} + z_2^{10} +z_3^{10} +z_4^5 +z_5^2 =
0,&(z_1:z_2:z_3:z_4:z_5) \in  \mP_{1,1,1,2,5}^4
\end{eqnarray}
$X_1$ can be described as triple cover of $\mP^3$ branched over a
sixtic, while $X_2$ can be 
described as a double cover of $\mP^3$ branched over an
octic~\cite{CMP101-341}. Maybe these d
escriptions will be useful for studying vector bundles on these
Calabi-Yau hypersurface by 
relating them to bundles over $\mP^3$, however, so far we have not been
able to achieve this. \\

In order to study the stringy geometry, we consider the mirror manifolds
$\widehat{X_i}$ given 
by the orbifold construction $\{p_i=0\}/G_i$
\begin{eqnarray}
p_1=&z_1^6 + z_2^6 +z_3^6 +z_4^6 +z_5^3 - 6\psi
z_1z_2z_3z_4z_5&G_1=\mZ_3\times\mZ_6^2\\
p_2=&z_1^8 + z_2^8 +z_3^8 +z_4^8 +z_5^2 - 8\psi
z_1z_2z_3z_4z_5&G_2=\mZ_2\times\mZ_8^2 \\
p_3=&z_1^{10} + z_2^{10} +z_3^{10} +z_4^5 +z_5^2 - 10\psi
z_1z_2z_3z_4z_5&G_3=\mZ_{10}^2  
\end{eqnarray}
The fundamental objects in our analysis are the periods of the
holomorphic 3-form 
$\widehat{\Omega}$ of $\widehat{X}$. We are interested in relating the
large volume limit point 
in the K\"ahler moduli space to the Gepner point, hence we have to
relate the periods at 
these two points as in~\cite{9906200}. At the Gepner point there is an
enhanced discrete 
symmetry which acts on the fundamental period $\varpi_0(\psi)$ by 
$\varpi_j(\psi)=A_G\varpi_0(\psi)=\varpi_0(\alpha^j\psi), j=0\dots k-1$,
where $\alpha^{k}=1$. 
Here $k=6,\,8,\,10$ for these three models~\cite{9205041}. Since
$b_3(X)=4$, we see that 
the periods $\varpi_j$ are not linearly independent. By considering
$\varpi_j$ for $|\psi|<1$ 
one obtains the following relations between the $\varpi_j$.
\begin{eqnarray}
\label{eq:relations1}
&\varpi_j + \varpi_{j+2} + \varpi_{j+4} = 0, \quad j = 0,1&\textrm{for
}\mP_{1,1,1,1,2}^4[6]\\
\label{eq:relations2}
&\varpi_j + \varpi_{j+4} = 0,\quad  j = 0,1,2,3&\textrm{for
}\mP_{1,1,1,1,4}^4[8]\\
\label{eq:relations3}
&\begin{array}{rcl}
\varpi_j + \varpi_{j+5} & = & 0 \quad j = 0,1,\dots,4 \\
\varpi_0 + \varpi_2 + \varpi_3 + \varpi_4 + \varpi_5& = & 0
\end{array}
&\textrm{for }\mP_{1,1,1,2,5}^4[10]
\end{eqnarray}
Following~\cite{9205041} we choose as period vectors 
$\varpi=(\varpi_2,\varpi_1,\varpi_0,\varpi_{k-1})^T,\, k=6,8,10$. On the
other hand, 
the large volume basis will be denoted by
$\Pi=(\Pi_6,\Pi_4,\Pi_2,\Pi_0)^T$. Then 
we have $\Pi = M\varpi$ where $M=KNm$ with $K$ as in~\cite{9906200} and
$m,N$ as 
in~\cite{9205041}. The change of basis for the three models is then, up
to an $Sp(4,\mZ)$ 
ambiguity,
\begin{eqnarray}
  \label{eq:M1}
  M = \left( 
  \begin{array}{cccc}
0&-1&1&0\\\noalign{\medskip}-1&0&3&2\\\noalign{\medskip}\textstyle\frac{1}{3}&
\textstyle\frac{1}{3}&-\textstyle\frac{1}{3}&-\textstyle\frac{1}{3}\\\noalign{\medskip}0&0&1&0
  \end{array}
  \right)
  &\textrm{for}&\mP_{1,1,1,1,2}^4[6]\\
  \label{eq:M2}
  M = \left( 
  \begin{array}{cccc}
0&-1&1&0\\\noalign{\medskip}-1&0&3&2\\\noalign{\medskip}\textstyle\frac{1}{2}&
\textstyle\frac{1}{2}&-\textstyle\frac{1}{2}&-\textstyle\frac{1}{2}\\\noalign{\medskip}0&0&1&0
  \end{array}
  \right)
  &\textrm{for}&\mP_{1,1,1,1,4}^4[8]\\
  \label{eq:M3}
  M = \left( 
  \begin{array}{cccc}
\hspace{1.8mm}0\hspace{1.8mm}&-1&\hspace{1.8mm}1\hspace{1.8mm}&0\\\noalign{\medskip}0&1&1&1\\\noalign{\medskip}1&0&0&-1\\\noalign{\medskip}0&0&1&0
  \end{array}
  \right)
  &\textrm{for}&\mP_{1,1,1,2,5}^4[10] 
\end{eqnarray}
With the help of these, the intersection form on $H^3(\widehat{X},\mZ)$
given in the 
large volume basis by ${\eta_L}_{14} = -{\eta_L}_{41} = -{\eta_L}_{23} =
{\eta_L}_{32} = 1$ 
can be transformed to the Gepner basis by
$\eta_G=M^{-1}\eta_L{M^{-1}}^T$
\begin{eqnarray}
  \eta_G = \left( 
  \begin{array}{cccc}
0&-1&2&0\\\noalign{\medskip}1&0&-1&2\\\noalign{\medskip}-2&1&0&-1\\\noalign{\medskip}0&-2&1&0
  \end{array}
  \right)
  &\textrm{for}&\mP_{1,1,1,1,2}^4[6]\\
  \eta_G = \left( 
  \begin{array}{cccc}
0&-1&2&-1\\\noalign{\medskip}1&0&-1&2\\\noalign{\medskip}-2&1&0&-1\\\noalign{\medskip}1&-2&1&0
  \end{array}
  \right) 
  &\textrm{for}&\mP_{1,1,1,1,4}^4[8]\\
  \eta_G = \left( 
  \begin{array}{cccc}
0&-1&1&1\\\noalign{\medskip}1&0&-1&1\\\noalign{\medskip}-1&1&0&-1\\\noalign{\medskip}-1&-1&1&0
  \end{array}
  \right) 
  &\textrm{for}&\mP_{1,1,1,2,5}^4[10] 
\end{eqnarray}
Using the relations~(\ref{eq:relations1}, (\ref{eq:relations2})
and~(\ref{eq:relations3}) we can 
express $\eta_G$ as a polynomial $I_G$ in the generator $g$ of the
enhanced discrete symmetry 
groups $\mZ_6$,$\mZ_8$ and $\mZ_{10}$
\begin{equation}
  \label{eq:IG1}
  \begin{array}{rclrl}
    I_G &=& -g + 2 g^2 - 2 g^4 + g^5&\textrm{for
}&\mP_{1,1,1,1,2}^4[6]\\
    I_G &=& -g + 2 g^2 - g^3 + g^5 -2 g^6 + g^7& \textrm{for
}&\mP_{1,1,1,1,4}^4[8]\\
    I_G &=& -g + g^2 + g^3 -2 g^4 + g^6 -g^7 -g^8  + 2 g^9 &\textrm{for
}&\mP_{1,1,1,2,5}^4[10]
  \end{array}
\end{equation}
We will return these models in section~\ref{sec:Gepner}.

\section{The geometry of the K3 fibrations}
\label{sec:two}
\setcounter{equation}{0}

We consider the models $\mP_{1,1,2,2,2}^4[8]$ which are given as degree
8 hypersurfaces $X$ in 
$\mP_{1,1,2,2,2}^4$ by e.g.
\begin{equation}
  \label{eq:degree8}
  z_1^8+z_2^8+z_3^4+z_4^4+z_5^4 = 0,\qquad (z_1:z_2:z_3:z_4:z_5) \in
\mP_{1,1,2,2,2}^4
\end{equation}
and the models $\mP_{1,1,2,2,6}^4[12]$ which are given as degree 12
hypersurfaces $X$ in 
$\mP_{1,1,2,2,6}^4$ by e.g.
\begin{equation}
  \label{eq:degree12}
  z_1^{12}+z_2^{12}+z_3^6+z_4^6+z_5^2 = 0,\qquad (z_1:z_2:z_3:z_4:z_5)
\in \mP_{1,1,2,2,6}^4
\end{equation}
They have been extensively studied in~\cite{9308083} and~\cite{9308122}.
Qualitative aspects of 
these models can and will be discussed together. The distinction between
them is only made where 
necessary. Both have a curve $C$ of $A_1$ singularities at $z_1=z_2=0$
which are of genus three 
and two, respectively. Blowing up these singularities gives an
exceptional divisor $E$ in $X$ 
which is a ruled surface. The degree one polynomials generate a linear
system $|L|$ which 
projects $X$ to $\mP^1$ with the fibers being $K3$ surfaces. The two
divisors $L$ and $E$ generate 
$H_4(X,\mZ)$. The degree two polynomials generate another linear system
$|H|$ which is related 
to the first by $|H| = |2L + E|$. The fiber of the ruled surface $E$
will be denoted by $l$. The 
intersection of two general members of $|H|$ and $|L|$ will be denoted
by $h$. The two classes 
$h$ and $l$ generate $H_2(X,\mZ)$. We choose the generators of the
complexified K\"ahler cone 
to be $(E,L)$, so that a generic K\"ahler class is written
$K=t_1E+t_2L$, where $(t_1,t_2)$ are 
classical coordinates on the K\"ahler moduli space of $X$.\\

These classes satisfy the following intersection relations. For
$\mP_{1,1,2,2,2}^4[8]$
\begin{equation}
  \label{eq:intersection8}
  \begin{array}{l}
    H^3=8,\quad H^2\cdot L=4,\quad H\cdot L^2 =0, \quad L^3=0,\\
    E^3=-16,\quad E^2\cdot L=4, \quad E\cdot L^2 = 0, \quad H\cdot
E\cdot L = 4\\
    h = \frac{1}{4} H\cdot L, \quad l=\frac{1}{4} H \cdot E\\
    L\cdot l = 1,\quad L\cdot h = 0, \quad H\cdot l = 0, \quad H\cdot h
= 1, 
\quad E \cdot l = -2, \quad E\cdot h = 1\\
    c_2(X)\cdot E =8,\quad c_2(X)\cdot H = 56 
  \end{array}
\end{equation}
and for $\mP_{1,1,2,2,6}^4[12]$
\begin{equation}
  \label{eq:intersection12}
  \begin{array}{l}
    H^3=4,\quad H^2\cdot L=2,\quad H\cdot L^2 =0, \quad L^3=0,\\
    E^3=-8,\quad E^2\cdot L=2, \quad E\cdot L^2 = 0, \quad H\cdot E\cdot
L = 2\\
    h = \frac{1}{2} H\cdot L, \quad l=\frac{1}{2} H \cdot E\\
    L\cdot l = 1,\quad L\cdot h = 0, \quad H\cdot l = 0, \quad H\cdot h
= 1, 
\quad E \cdot l = -2, \quad E\cdot h = 1\\
    c_2(X)\cdot E =4,\quad c_2(X)\cdot H = 52 
  \end{array}
\end{equation}

When we include $\alpha'$ corrections to the classical geometry, we do
this by considering 
the complex structure moduli space of the mirror $\widehat{X}$ which
does not get any such 
corrections. The mirror family $\widehat{X}$ is given by $\{p=0\}/G$
where for 
$\mP_{1,1,2,2,2}^4[8]$ $G=\mZ_4^3$ and 
\begin{equation}
  \label{eq:mirror8}
  p = z_1^8+z_2^8+z_3^4+z_4^4+z_5^4 -8\psi z_1z_2z_3z_4z_5 -2\phi
z_1^4z_2^4
\end{equation}
while for $\mP_{1,1,2,2,6}^4[12]$ $G=\mZ_6^2\times\mZ_2$ and
\begin{equation}
  \label{eq:mirror12}
  p = z_1^{12}+z_2^{12}+z_3^6+z_4^6+z_5^2 -12\psi z_1z_2z_3z_4z_5 -2\phi
z_1^6z_2^6
\end{equation}
$\psi$ and $\phi$ parametrize the moduli space of complex structures of
the mirror $\widehat{X}$.

On the K\"ahler side, the prepotential $\cF$ determines the periods
\linebreak
$\Pi=(\cF^0,\cF^1,\cF^2,1,t_1,t_2)^T$ of the holomorphic 3-form
$\widehat{\Omega}$ on 
$\widehat{X}$, where $\cF^i=\frac{\partial \cF}{\partial t_i},\;i=1,2$
and $\cF^0=2\cF-t_i\cF^i$. 
In our two cases we have for $\mP_{1,1,2,2,2}^4[8]$
(see~\cite{9308083})  
\begin{equation}
  \label{eq:prepotential8}
  \cF = -\textstyle\frac{4}{3}t_1^3 - 2t_1^2t_2 +
\textstyle\frac{7}{3}t_1 +t_2 +\textrm{const}
\end{equation}
and for $\mP_{1,1,2,2,6}^4[12]$\footnote{see
appendix~\ref{sec:Candelas}}
\begin{equation}
  \label{eq:prepotential12}
  \cF = -\textstyle\frac{2}{3}t_1^3 - t_1^2t_2 +
\textstyle\frac{13}{6}t_1 +t_2 +\textrm{const}
\end{equation}
Using the intersection numbers~(\ref{eq:intersection8})
and~(\ref{eq:intersection12}) these can 
be combined to~\cite{9406055}
\begin{equation}
  \label{eq:prepotential}
  \cF = -\frac{1}{3!}\left(H^3\, t_1^3 +3H^2\cdot L\, t_1^2t_2\right)
+\frac{t_1}{24}\int_{\widehat{X}} 
c_2(\widehat{X})H  +\frac{t_2}{24}\int_{\widehat{X}} c_2(\widehat{X})L
-\frac{\zeta(3)i}{2(2\pi)^3}\chi(\widehat{X})
\end{equation}
As in~\cite{9910172} the choice of $(E,L)$ as the generators of the
complexified K\"ahler cone 
leads to a non-canonically symplectic intersection form on
$H^3(\widehat{X},\mZ)$ in the large 
volume limit
\begin{equation}
  \label{eq:intersectionL}
  \eta_L =\left(\begin{array}{cccccc}
\hspace{1.8mm}0\hspace{1.8mm}&0&\hspace{1.8mm}0\hspace{1.8mm}&-1&0&0\\\noalign{\medskip}0&0&0&0&-1&2\\\noalign{\medskip}0&0&0&0&0&-1
\\\noalign{\medskip}1&0&0&0&0&0
\\\noalign{\medskip}0&1&0&0&0&0\\\noalign{\medskip}0&-2&1&0&0&0
  \end{array}\right)
\end{equation}
Hence the period vector in the $(E,L)$ basis becomes $\Pi=(\cF^0,\cF^1-
2\cF^2,\cF^2,1,t_1,t_2)^T$. 
Next, we have to relate this period vector to the one at the Gepner
point 
$\varpi=(\varpi_0,\dots,\varpi_5)^T$ by $\Pi = M\varpi$. Due to the
enhanced symmetry, 
$\mZ_k$, where $k$ is the degree of the hypersurface, at this point, the
periods $\varpi_j, j=0\dots k-1$ are 
not all linearly independent. They satisfy a set of relations which
allow us to restrict them to 
$h^3=6$ of these periods. These relations are for $\mP_{1,1,2,2,2}^4[8]$
\begin{equation}
  \label{eq:relations8}
  \varpi_j + \varpi_{j+2} + \varpi_{j+4} + \varpi_{j+6}  =  0 \qquad j =
0,1
\end{equation}
and for $\mP_{1,1,2,2,6}^4[12]$
\begin{equation}
  \label{eq:relations12}
  \varpi_j + \varpi_{j+6} = 0 \qquad j=0,1,\dots,5
\end{equation}
The basis transformation matrix $m$ from the Gepner point to the large
volume limit point can 
be obtained by analytic continuation as in~\cite{9308083}\footnote{See
also appendix~\ref{sec:Candelas}.}.
\begin{eqnarray}
  \label{eq:M4}
  M = \left( 
  \begin{array}{cccccc}
-1&\hspace{1.6mm}1\hspace{1.6mm}&0&\hspace{1.6mm}0\hspace{1.6mm}&0&0\\\noalign{\medskip}-\frac{1}{2}&\frac{3}{2}&-2&0&-\frac{1}{2}&-\frac{1}{2}
\\\noalign{\medskip}1&0&1&0&0&0\\\noalign{\medskip}1&0&0&0&0&0\\\noalign{\medskip}-
\textstyle\frac{1}{4}&0&\textstyle\frac{1}{2}&0&\textstyle\frac{1}{4}&0\\\noalign{\medskip}
\textstyle\frac{1}{4}&\textstyle\frac{3}{4}&-\textstyle\frac{1}
{2}&\textstyle\frac{1}{2}&-\textstyle\frac{1}{4}&\textstyle\frac{1}{4}
  \end{array}
  \right) &\textrm{for}&\mP_{1,1,2,2,2}^4[8]\\
  \label{eq:M5}
   M  = \left( 
  \begin{array}{cccccc}
-1&\hspace{1.6mm}1\hspace{1.6mm}&0&\hspace{1.6mm}0\hspace{1.6mm}&0&0\\\noalign{\medskip}-\frac{1}{2}&\frac{3}{2}&-\frac{3}{2}&\frac{1}{2}&
-\frac{1}{2}&-\frac{1}{2}\\\noalign{\medskip}1&0&1&0&0&0\\\noalign{\medskip}1&0&0&0&0&0
\\\noalign{\medskip}-\textstyle\frac{1}{2}&0&\textstyle\frac{1}{2}&0&\textstyle\frac{1}{2}&0
\\\noalign{\medskip}\textstyle\frac{1}{2}&\textstyle\frac{1}{2}&-\textstyle\frac{1}{2}&
\textstyle\frac{1}{2}&-\textstyle\frac{1}{2}&\textstyle\frac{1}{2}
  \end{array}
  \right) &\textrm{for}&\mP_{1,1,2,2,6}^4[12]
\end{eqnarray}
Now we can express the intersection matrix in the basis of the periods
at the Gepner 
point by $\eta_G= M^{-1} \eta_L {M^{-1}}^T$. 
\begin{eqnarray}
  \label{eq:matrixetaG}
  \eta_G = \left( 
  \begin{array}{cccccc}
0&1&0&-3&0&3\\\noalign{\medskip}-1&0&1&0&-3&0\\\noalign{\medskip}0&-1&0&1&0&-3
\\\noalign{\medskip}3&0&-1&0&1&0
\\\noalign{\medskip}0&3&0&-1&0&1\\\noalign{\medskip}-3&0&3&0&-1&0
  \end{array}
  \right)
  &\textrm{for}&\mP_{1,1,2,2,2}^4[8]\\
 \eta_G = \left( 
  \begin{array}{cccccc}
0&1&0&-2&0&1\\\noalign{\medskip}-1&0&1&0&-2&0\\\noalign{\medskip}0&-1&0&1&0&-2
\\\noalign{\medskip}2&0&-1&0&1&0
\\\noalign{\medskip}0&2&0&-1&0&1\\\noalign{\medskip}-1&0&2&0&-1&0
  \end{array}
  \right)
  &\textrm{for}&\mP_{1,1,2,2,6}^4[12]
\end{eqnarray}
Using the relations~(\ref{eq:relations8}) and~(\ref{eq:relations12})
respectively 
we can express $\eta_G$ as a polynomial $I_G$ in the generator $g$ of
the enhanced discrete 
symmetry groups $\mZ_8$ and $\mZ_{12}$.
\begin{equation}
  \label{eq:IG2}
  \begin{array}{rclrl}
    I_G&=&g - 3 g^3 +3 g^5 -g^7&\textrm{for}&\mP_{1,1,2,2,2}^4[8]\\
    I_G&=&g - 2g^3 + g^5 - g^7 + 2g^9 -
g^{11}&\textrm{for}&\mP_{1,1,2,2,6}^4[12] 
  \end{array}
\end{equation}
We will return to these expressions in section~\ref{sec:Gepner}.

\section{D-branes and periods}
\label{sec:periods}
\setcounter{equation}{0}
This section follows closely~\cite{9910172} to which we refer for
further details. Let 
 $n=(n_6,n_4^1,n_4^2,n_0,n_2^1,n_2^2)$ be an integral vector of
$H^3(\widehat{X},\mZ)$ which 
describes the low energy charges of the D-brane. Then the central charge
is
\begin{equation}
Z(n) = n\cdot\Pi = n_6\Pi^1 + n_4^1\Pi^2 + n_4^2\Pi^3 + n_0\Pi^4 +
n_2^1\Pi^5 + n_2^2\Pi^6
\end{equation}
We want to map these charges to the topological invariants of the
corresponding K-theory 
class $\xi$ which are given by the Chern character $\chern(\xi)$. This
is done using the 
exact form of D-brane Chern-Simons couplings
\begin{equation}
\label{eq:CS}
Q = \chern(\xi)\sqrt{\todd(X)} \in H^{\mbox{\scriptsize even}}(X,\mZ)
\end{equation}
The central charge is
\begin{equation}
\label{eq:Z}
Z(K') = \int_X \frac{{K'}^3Q^0}{6} - \frac{{K'}^2\cdot Q^2}{2} + K'\cdot
Q^4 - Q^6
\end{equation}
where $K' = t_1H + t_2L$ is the generic K\"ahler class before the change
of basis to $(E,L)$. 
We apply these general ideas to two cases of $D$-brane systems. The
first are those with 
nonzero $D6$-brane charge which can be represented by holomorphic vector
bundles (more precisely 
coherent sheaves~\cite{9609017}) $V$ on $X$. In order for the
corresponding $D$-brane 
configuration to be supersymmetric $V$ must be stable~\cite{9804131}.
From~(\ref{eq:CS}) we obtain
\begin{equation}
Q=\left(r,\ch_1(V),\chern_2(V)+\frac{r}{24}\ch_2(X),\chern_3(V)+\frac{1}{24}\ch_1(V)\ch_2(X)\right)
\end{equation}
and from~(\ref{eq:Z})
\begin{equation}
  Z(Q) = \frac{r}{6} {K'}^3 -\frac{1}{2}\chern_1(V)\cdot{K'}^2 + 
\left(\chern_2(V)+\frac{r}{24}\ch_2(X)\right){K'} -
\left(\chern_3(V)+\frac{1}{24}\ch_1(V)\ch_2(X)\right) 
\end{equation}
By comparison one obtains for the Chern classes of $V$ on both models
\begin{eqnarray}
  r(V)&=&n_6\\
  \ch_1(V)&=&n_4^1E+n_4^2L\\
  \ch_2(V)&=&\left(4n_4^1(n_4^2-n_4^1)-n_2^1\right)h
+\left(2(n_4^1)^2-n_2^2\right)l\\
  \begin{array}{r}\ch_3(V)\\ \\ \end{array}&\begin{array}{c}=\\ \\
\end{array}&\begin{array}{l}
2\left(n_4^1\right)^2\left(-4n_4^1+3n_4^2\right)+3n_4^1\left(2n_2^2-n_2^1\right)-3n_4^2n_2^2\\
    -6n_0-12n_4^2+\chi_C n_4^1
  \end{array}
\end{eqnarray}
where $\chi_C=-4$ for $\mP_{1,1,2,2,2}^4[8]$ and $\chi_C=-2$ for
$\mP_{1,1,2,2,6}^4[12]$.\\

The second case consists of systems of $D4$-branes on the Calabi-Yau
manifolds wrapped on holomorphic 
submanifolds $i:D\hookrightarrow X$ where $D\in H_4(X,\mZ)$ as has been
shown in~\cite{9507158} 
and~\cite{9606112}. Using the Riemann-Roch-Grothendieck theorem, the
central charge associated to 
the vector $Q$
\begin{equation}
Q=\left(0,rD,i_*\ch_1(V)+\frac{r}{2}i_*\ch_1(D),\chern_2(V)+\frac{1}{2}\ch_1(V)\ch_1(D)+
\frac{r}{8}\ch_1(D)^2+\frac{r}{24}\ch_2(D)\right)
\end{equation}
becomes~\cite{9910172}
\begin{equation}
  \label{eq:1}
  \begin{array}{rcl}
  Z(Q)&=&-\frac{r}{2}{K'}^2\cdot
D+\left(i_*\ch_1(V)+\frac{r}{2}i_*\ch_1(D)\right)K'\\
      &
&-\chern_2(V)-\frac{1}{2}\ch_1(V)\ch_1(D)-\frac{r}{8}\ch_1(D)^2-\frac{r}{24}\ch_2(D)
  \end{array}
\end{equation}
$D4$-branes wrapped on the exceptional divisor $E$ correspond to BPS
states with charge vectors 
$n=(0,n_4^1,0,n_0,n_2^1,n_2^2)$ with central charge
\begin{equation}
  Z(n)=n_4^1\cF^1 + n_2^1t_1 + n_2^2t_2 +n_0
\end{equation}
Using $c_2(E) = 2\chi_C$ the Chern classes of $V$ can be expressed as
\begin{eqnarray}
  r(V)&=&n_4^1\\
  \ch_1(V)&=&\left(n_2^1+\chi_Cn_4^1\right)h
+\left(n_2^2-\textstyle\frac{\chi_C}{2} n_4^1\right)l\\
  \chern_2(V)&=&-\frac{3}{2}\chi_Cn_4^1-\frac{1}{2}n_2^1+n_2^2-n_0  
\end{eqnarray}
The $D4$-branes wrapped on the $K3$ fiber $L$ correspond to BPS states
with charge vectors 
$n=(0,0,n_4^2,n_0,n_2^1,0)$ with central charge\footnote{$n_2^2=0$
follows e.g. from 
consistency between~(\ref{eq:1}) and~(\ref{eq:2})}
\begin{equation}
  \label{eq:2}
  Z(n)=n_4^2\cF^2 + n_2^1t_1 +n_0
\end{equation}
The Chern classes of $V$ can be expressed as
\begin{eqnarray}
  r(V)&=&n_4^2\\
  \ch_1(V)&=&n_2^1h\\
  \chern_2(V)&=&-2n_4^2-n_0
\end{eqnarray}
This gives for the Mukai vector for $K3$ fibers $v(V)$~\cite{9609017}
\begin{equation}
  v(V)=\left(r(V),c_1(V),r(V)+\chern_2(V)\right) =
\left(n_4^2,n_2^1h,-n_4^2-n_0\right)\in 
H^{\mbox{\scriptsize even}}(K3,\mZ)
\end{equation}
There is a natural inner product on the space of Mukai
vectors~\cite{9906200}
\begin{equation}
  \langle v,v'\rangle = \langle(r,s,\ell),(r',s',\ell')\rangle = s\cdot
s' - r \cdot 
\ell' -\ell\cdot r'
\end{equation}
Applied to our vector $v(V)$ this gives
\begin{equation}
  \label{eq:Mukai}
  \langle v,v\rangle =
2n_4^2(n_4^2+n_0)-\frac{1}{\chi_C}\left(n_2^1\right)^2
\end{equation}
where the factor $-\frac{1}{\chi_C}$ arises as follows: 
$h\cdot h|_{L}=\frac{1}{{\chi_C}^2}H\cdot
H|_{L}=\frac{1}{{\chi_C}^2}H^2\cdot L$. 
A theorem of Mukai~\cite{SugakuExp1-139} shows that the space of
coherent simple 
semistable sheaves with Chern classes specified by $Q$ is smooth and
compact and 
has complex dimension
\begin{equation}
  d(n)= \langle v,v\rangle +2
\end{equation}
This will be used when making the correspondence between supersymmetric
D-brane configurations 
at the large volume limit and the boundary states in the Gepner model to
which we now turn.

\section{Boundary states in the Gepner model}
\label{sec:Gepner}
\setcounter{equation}{0}
The Gepner point is characterized by its enhanced discrete (quantum)
symmetry and hence by the 
fact that the corresponding superconformal field theory is exactly
solvable. The corresponding 
boundary conformal field theory has been solved for the rational
boundary states in~\cite{9712186} 
and studied further in~\cite{9808080} and~\cite{9902120}. Here we want
to extend the analysis 
of~\cite{9906200} and~\cite{9910172} to the models described in the
previous sections. 
In particular, we will compute the symplectic intersection form on the
BPS charge lattice in 
the superconformal field theory.\\

The Gepner model $(k_1,k_2,k_3,k_4,k_5)$ is given by the tensor product
of $r=5$ minimal 
models at level $k_j$ subject to a projection onto states with odd
integer $U(1)$ charges and 
addition of ``twisted'' sectors in order to keep the theory modular
invariant. We include a 
$k=0$ factor if present. The superconformal primaries of the minimal
models are labelled by 
3 integers, $(l_j,m_j,s_j)$ with
\begin{equation}
  0\leq l_j\leq k_j,\quad |m_j-s_j|\leq l_j, \quad s_j\in\{-1,0,1,2\},
\quad l_j+m_j+s_j=0 \mod 2
\end{equation}
We also introduce the vectors $\lambda=(l_1,\dots,l_r)$ for the $l_j$
quantum number, 
$\mu=(s_0;m_1,\dots,m_r;s_1,\dots,s_r)$ for the charges and spin
structures. The rational 
boundary states are constructed by considering each factor separately
and then subjecting 
it to Cardy's consistency condition for modular invariance. They are
labelled by 
$\alpha=(L_j,M_j,S_j)$ and an automorphism $\Omega$ of the chiral
symmetry algebra giving 
either A- or B-type boundary conditions. They are
\begin{equation}
  |\alpha\rangle\rangle =
\frac{1}{\kappa_\alpha^\Omega}\sum_{\lambda,\mu}\delta_\beta
\delta_\Omega B_\alpha^{\lambda,\mu}|\lambda,\mu\rangle\rangle_\Omega
\end{equation}
The states $|\alpha\rangle\rangle$ are Cardy states while
$|\lambda,\mu\rangle\rangle_\Omega$ 
are Ishibashi states~\cite{MPLA4-251}. $\kappa_\alpha^\Omega$ and
$B_\alpha^{\lambda,\mu}$ are 
given in~\cite{9712186} and~\cite{9906200}. $\delta_\beta$ is a
Kronecker delta function 
enforcing both odd $U(1)$ integral charge and the condition that all
factors of the tensor 
product have the same spin structure. For the B-type boundary states it
implies that the 
physically inequivalent choices for $M_j$ can be described by the
quantity
\begin{equation}
  M=\sum_{j=1}^r\frac{K'M_j}{2k_j+4}
\end{equation}
where $K'=\textrm{lcm}\{2k_j+4\}$. Hence, we will label the B-type
boundary states by 
$|L_1,\dots,L_r;M;S\rangle\rangle_B$. Due to the symmetry of the
superconformal primaries 
of the minimal models
$\chi_{m_j,s_j}^{l_j}=\chi_{m_j+k_j+2,s_j+2}^{k_j-l_j}$ one can 
restrict the values of the $L_j$ to $0\leq L_j \leq \lfloor
\frac{k_j}{2}\rfloor$. The 
delta function $\delta_\Omega$ guarantees that the
$|\lambda,\mu\rangle\rangle_\Omega$ 
appear in the closed string partition function. While not giving any
condition for 
the A-type boundary states, it requires that $m_j = b_j\mod k_j+2$ for
some $b_j$. We will 
denote the set of states which is obtained from a given state
$\BS{L_1,\dots,L_r;M;0}$ by 
applying to it the generator of the quantum symmetry as its $L$-orbit.\\

The properties of the Gepner models corresponding to the Calabi-Yau
spaces in the preceding 
sections are summarized in the follwing table:
\begin{equation}
  \begin{array}{|c|c|c|}
    \hline
    \textrm{CY family} &\textrm{Gepner model} & \textrm{Symmetry
group}\\
    \hline
    \mP_{1,1,1,1,2}^4[6]& (4,4,4,4,1) &
\frac{\mZ_6^4\times\mZ_3}{\mZ_6}\times\mZ_2\\
    \hline
    \mP_{1,1,1,1,4}^4[8]& (6,6,6,6,0)
&\frac{\mZ_8^4\times\mZ_2}{\mZ_8}\times\mZ_2\\
    \hline
    \mP_{1,1,1,2,5}^4[10]& (8,8,8,3,0)
&\frac{\mZ_{10}^3\times\mZ_5\times\mZ_2}{\mZ_{10}}\times\mZ_2\\
    \hline
    \mP_{1,1,2,2,2}^4[8]& (6,6,2,2,2)
&\frac{\mZ_8^2\times\mZ_4^3}{\mZ_8}\times\mZ_2\\
    \hline
    \mP_{1,1,2,2,6}^4[12]& (10,10,4,4,0) &
\frac{\mZ_{12}^2\times\mZ_6^2\times\mZ_2}{\mZ_{12}}\times\mZ_2\\
    \hline
  \end{array}
\end{equation}

In order to find the geometric interpretation of the Gepner model
boundary states 
we consider the intersection form of their charge lattice. In the
conformal field 
theory this can be computed by the Witten index $I_\Omega=\tr_R(-1)^F$
in the open 
string sector, as was explained in~\cite{9610236} and~\cite{9903031}.
This is a 
topological invariant and hence is unaffected by marginal deformations
of the SCFT. 
In~\cite{9906200} it has been related to the index of the Dirac operator
on $V^*\otimes W$, 
where $V$ and $W$ are the vector bundles on the intersecting branes. For
the A-type boundary 
states it is given by~\cite{9906200}
\begin{equation}
I_A=\frac{1}{C}(-1)^{\frac{S-\widetilde{S}}{2}}\sum_{\nu_0=0}^{K-1}\prod_{j=1}^r
N_{L_j,\widetilde{L}_j}^{2\nu_0+M_j-\widetilde{M}_j}
\end{equation}
and for the B-type boundary states by
\begin{equation}
  \label{eq:IB}
  I_B=\frac{1}{C}(-1)^{\frac{S-\widetilde{S}}{2}}\sum_{\{m_a\}}
\delta_{\frac{M+\widetilde{M}}{2}+\sum_{j=1}^r\frac{K'}{2k_j+4}(m_j+1)}^{(K')}
\prod_{j=1}^rN_{L_j,\widetilde{L}_j}^{m_j-1}
\end{equation}
where $K=\textrm{lcm}\{4,2k_j+4\}$ and $N_{L,\widetilde{L}}^l$ are the
extended 
$SU(2)_k$ fusion coefficients~\cite{9906200}.\\

Let us first consider the A-type boundary states with $L_j=0$. The next
table shows 
the results for our models
\begin{equation}
  \begin{array}{|c|c|c|}
    \hline
    \textrm{Gepner model} & I_A & \rank(I_A) \\
    \hline
    (4,4,4,4,1) & (1-g_2g_3g_4g_5)(1-g_2^5)(1-g_3^5)(1-g_4^5)(1-g_5^2) &
208\\
    \hline
    (6,6,6,6,0) & (1-g_2g_3g_4g_5)(1-g_2^7)(1-g_3^7)(1-g_4^7)(1-g_5) &
300 \\
    \hline
    (8,8,8,3,0) & (1-g_2g_3g_4g_5)(1-g_2^9)(1-g_3^9)(1-g_4^4)(1-g_5^2) &
292\\
    \hline
    (6,6,2,2,2) & (1-g_2g_3g_4g_5)(1-g_2^7)(1-g_3^3)(1-g_4^3)(1-g_5^3) &
168\\
    \hline
    (10,10,4,4,0) &
(1-g_2g_3g_4g_5)(1-g_2^{11})(1-g_3^5)(1-g_4^5)(1-g_5) & 254\\
    \hline
    (16,16,16,1,0) &
(1-g_2g_3g_4g_5)(1-g_2^{17})(1-g_3^{17})(1-g_4^2)(1-g_5) & 272 \\
    \hline
  \end{array}
\end{equation}
where $g_i,\,i=2..5$ are the generators of the symmetry group satisfying
$g_i^{k_i+2}=1$. 
For completeness we have included the elliptic fibration
$\mP_{1,1,1,6,9}^4[18]$ considered 
in~\cite{9910172}. It turns out that the $\rank(I_A)$ can be related to
a geometric quantity
\begin{equation}
  \rank(I_A) = \tilde{b}_3(X)
\end{equation}
where $\tilde{b}_3(X)$ denotes the third Betti number of the
corresponding Calabi-Yau family 
without the contributions form non-polynomial deformations of the
complex structure~\cite{9308122}.
It can be checked that this holds for any Fermat hypersurface $X$
irrespective of $h^{1,1}(X)$.
This means that the rank of this intersection matrix counts the number
of independent 4-cycles 
on the mirror Calabi-Yau $\widehat{X}$ except those which are coming
from a non-toric blow-up. 
For the two $K3$ fibrations under consideration the number of non-toric
4-cycles is given by 
$g = 1-\frac{\chi_C}{2}$~\cite{9601108} with the $\chi_C$ being
introduced in the previous 
section. It has been recently shown~\cite{9912151} that these complex
structure deformations 
can lead to a superpotential in the non-compact space-time. \\

Next we consider the B-type boundary states. Recall that these states
are described by the 
single integer $M\in\mZ_{K'}$ and that the $g_j$ for different $j$ are
identified. First, 
we are looking again at the $L_j=0$ states. In this case, the
intersection matrix $I_B$ can 
be related to the intersection polynomial $I_G$ from the previous
sections.
\begin{equation}
  \begin{array}{|c|c|c|}
    \hline
    \textrm{Gepner model} & I_B & \textrm{Quantum symmetry}\\
    \hline
    (4,4,4,4,1) & (1-g^5)^4(1-g^4) & \mZ_6\\
    \hline
    (6,6,6,6,0) & (1-g^7)^4(1-g^4) &\mZ_8\\
    \hline
    (8,8,8,3,0) & (1-g^9)^3(1-g^8)(1-g^5) & \mZ_{10}\\
    \hline
    (6,6,2,2,2) & (1-g^7)^2(1-g^6)^3 & \mZ_8\\
    \hline
    (10,10,4,4,0) & (1-g^{11})^2(1-g^{10})^2(1-g^6) &\mZ_{12}\\
    \hline
  \end{array}
\end{equation}
It turns out that in all the examples considered, including the quintic
in $\mP^4$ and the 
elliptic fibration $\mP_{1,1,1,6,9}^4[18]$, the relation between the
intersection matrix $I_B$ 
calculated from conformal field theory and the intersection matrix $I_G$
found in~(\ref{eq:IG1}) 
and~(\ref{eq:IG2}) by using the mirror map at the large volume limit is
the same
\begin{equation}
  I_B = (1-g)\;I_G\;(1-g^{-1})
\end{equation}
A convenient way to find the charges for $L_j\not = 0$ has been found
in~\cite{9910172}. 
One replaces each factor $N_{L_j,\widetilde{L}_j}^{m_j-1}$ by a factor
$n_{L,\widetilde{L}}$. 
Starting from $n_{0,0}=(1-g^{-1})$ one applies to it the linear
transformation
\begin{equation}
  t_L={t_L}^T=\sum_{l=-\frac{L}{2}}^{\frac{L}{2}}g^l 
\end{equation}
to obtain $n_{L,\widetilde{L}}=t_Ln_{0,0}{t_{\widetilde{L}}}^T$. The
charge of the boundary 
state $q_B$ in the Gepner basis is then given by
$q_Bt_{L_1}t_{L_2}t_{L_3}t_{L_4}t_{L_5}$. In 
order to obtain the charges at the large volume limit we substitute the
matrix $A_L$ for $g$, 
where $A_L$ is given for the different models under discussion in the
appendix~\ref{sec:AL}. \\

Before we start looking at the spectrum of B-type boundary states we
will mention how to compute 
the number of boundary marginal operators. For a single boundary state
$L=|\,L_j;M;S\rangle\rangle_B$ 
it is the constant term in~\cite{9910172}  
\begin{equation}
  P_B =
\frac{1}{2}\widetilde{n}_{L_1,L_1}\widetilde{n}_{L_2,L_2}\widetilde{n}_{L_3,L_3}
\widetilde{n}_{L_4,L_4}\widetilde{n}_{L_5,L_5}-\rho
\end{equation}
where $\widetilde{n}_{L_j,L_j}=|n_{L_j,L_j}|$ and $\rho=2^{\gamma-1}$.
$\gamma$ counts the number 
of $L_j$ which are equal to $\frac{k_j}{2}$. This is due to the symmetry
$\BS{L_j}=\BS{k_j-L_j}$ 
in a $L_j$-orbit which halves the number of states except for $L_j =
\frac{k_j}{2}$. In the 
expression for $\rho$, $k_j=0$ factors are taken into account.

\section{Boundary states and D-branes}
\label{sec:D-branes}
\setcounter{equation}{0}
In this section we put everything together and establish an explicit
correspondence between the 
D-branes described by rational boundary states in the Gepner model and
supersymmetric D-brane 
configurations on the Calabi-Yau spaces discussed above. \\
 
For the 1-parameter models considered here we are faced with the same
problem as for the 
quintic: there is a general lack of knowledge of vector bundles on these
spaces. Hence we 
can only consider some particular aspects. In~\cite{9906200} it was
noticed that there was 
no D0 brane in the spectrum of the quintic at the Gepner point. Even
though this only means 
that there is no corresponding rational boundary state for the D0-brane,
it was argued later 
on in~\cite{9910170} that there might be a line of marginal stability
which might prevent 
its existence at the Gepner point at all. However, this seems not to be
generically the 
case, as has been shown in~\cite{9910172} and as we will show here. Let
us discuss some 
interesting states in our models. In $\mP_{1,1,1,1,2}[6]$ we do not find
the D0-brane either, 
however there is an interesting state in the orbit of
$L=|\,1,1,0,0,0\rangle\rangle_B$ which 
is a D2-brane with charge $Q_2=3$. This state has 8 marginal operators
in the Gepner model.\\

For $\mP_{1,1,1,1,4}[8]$ some interesting boundary states are listed in
the following table 
(there are many more boundary states which, however, we do not discuss
here)
\begin{equation}
  \begin{array}{|c|c|c|c|}
    \hline
     
\textrm{$L$-orbit}&\textrm{charge}&\textrm{\#moduli}&\textrm{\#vacua}\\
    \hline
      \BS{1,1,0,0,0}&(0,0,2,0)&7&1\\
    \hline
      \BS{2,0,0,0,0}&(1,0,0,-2)&6&1\\
    \hline
      \BS{2,1,0,0,0}&(0,1,2,0)&11&1\\
    \hline
      \BS{3,0,0,0,0}&(0,0,0,2)&6&2\\
    \hline
  \end{array}
\end{equation}
Here we find the $D0$-brane in the last line. Since the number of vacua
is two, we might 
think of the boundary state as two different $D0$-branes, each of which
would have 3 moduli. 
This matches with the expectation that the moduli space of the
$D0$-brane is the whole 
Calabi-Yau manifold. As was argued in~\cite{9910172} this might be a
sign of a Coulomb branch 
in the worldvolume theory in which the gauge group is $U(1)^2$. In the
first line we also find 
the $D2$-brane wrapping some 2-cycle while there was no pure $D4$-brane
found. Particularly 
interesting are the states in the second and third line. They describe
$D6-\overline{D0}$ 
and $D4-D2$ bound states respectively. A $D4-D2$ system is known to
break supersymmetry 
completely in flat space. The $D4-D2$ potential which is difficult to
determine on a curved 
space is expected to approach the flat space result in the large volume
limit. Hence this 
configuration should become non-supersymmetric and repulsive for a
sufficiently big radius 
which prevents in particular the formation of a bound state. This gives
a further example 
of a supersymmetric boundary state at the Gepner point decaying into a
non-supersymmetric 
combination of D-branes at large volume point. This phenomenon was first
observed in~\cite{9910172}.\\

Next, we consider some important boundary states for
$\mP_{1,1,1,2,5}[10]$ which are listed below.
\begin{equation}
  \begin{array}{|c|c|c|c|}
    \hline
     
\textrm{$L$-orbit}&\textrm{charge}&\textrm{\#moduli}&\textrm{\#vacua}\\
    \hline
      \BS{0,0,0,1,0}&(0,0,0,1)&3&1\\ 
    \hline
      \BS{2,0,0,0,0}&(0,0,2,0)&4&1\\
    \hline
      \BS{2,0,0,1,0}&(0,1,0,0)&8&1\\
    \hline
      \BS{2,0,0,1,0}&(0,1,1,0)&8&1\\
    \hline
      \BS{4,0,0,0,0}&(0,0,0,2)&6&2\\
    \hline
  \end{array}
\end{equation}
Again we find the $D0$-brane in the spectrum of the boundary states. In
this model it appears in 
two different ways: Once as a single $D0$-brane in the first line with
the expected dimension of 
the moduli space and once as two different $D0$-branes in the last line,
presicely in the same 
way as in the case above. Furthermore we find in addition to the
$D2$-brane wrapping some 2-cycle 
in the second line also a $D4$-brane wrapping some 4-cycle. Finally
there is a supersymmetric 
bound state of a $D4-D2$ system at the Gepner point corresponding to a
non-supersymmetric 
configuration at the large volume point.\\

For the remainder of this section we turn to the more interesting
$K3$-fibrations. The following 
table gives all the boundary states which describe brane configurations
wrapped on the $K3$ fiber. 
The criterion is that their charge vector is of the general form
$(0,0,n_4^2,n_0,n_2^1,0)$. 
The anti-branes whose charge vector has the opposite overall sign are
not given in the table. 
\begin{equation}
  \begin{array}{|c|c|c|c|c|}
    \hline
      \textrm{$L$-orbit}&\textrm{Mukai vector
$v=(n_4^2,n_2^1,-n_4^2-n_0)$}&
\textrm{\#moduli}&\textrm{\#vacua}&\textrm{$d$}\\
    \hline
      \BS{1,0,0,0,0}& (2,-2,1)\;(1,0,1)\;(1,-2,2)&1&1&0\\
    \hline
      \BS{3,0,0,0,0}& (1,0,-1)\;(0,2,-1)\;(1,-2,0)&5&1&4\\
    \hline
      \BS{3,0,1,0,0}& (1,-4,1)\;(2,-2,-1)\;(1,2,-2)&9&1&8\\
    \hline
      \BS{3,0,1,1,0}& (0,6,-3)\;(3,-6,0)\;(3,0,-3)&21&1&20\\
    \hline
      \BS{5,0,0,0,0}& (2,0,0)\;(0,0,2)\;(2,-4,2)&6&2&2\\
    \hline
      \BS{5,0,1,0,0}& (2,-4,0)\;(2,0,-2)\;(0,4,-2)&14&2&10\\
    \hline
      \BS{5,0,1,1,0}& (2,-8,2)\;(4,-4,-2)\;(2,4,-4)&30&2&26\\
    \hline
      \BS{5,0,2,0,0}& (0,4,0)\;(4,-4,0)\;(0,4,-4)&20&4&10\\
    \hline
      \BS{5,0,2,1,0}& (4,-8,0)\;(4,0,-4)\;(0,8,-4)&44&4&34\\
    \hline
      \BS{5,0,2,2,0}& (4,-12,4)\;(4,-4,-4)\;(4,4,-4)&64&8&40\\
    \hline
  \end{array}
\end{equation}
The BPS condition for the charges is~\cite{9609017} $d\geq 0$ with $r>0$
or 
$r(V)=0,\ch_1(V)>0$ or $r=\ch_1=0,\chern_2(V)<0$. This is satisfied for
all the 
charges above. Note that all the states in a given orbit lead to the
same dimension 
$d$ which provides a check on~(\ref{eq:Mukai}). 

There are some further boundary states of particular interest listed
below
\begin{equation}
  \begin{array}{|c|c|c|c|c|}
    \hline
     
\textrm{$L$-orbit}&\textrm{charges}&\textrm{\#moduli}&\textrm{\#vacua}\\
    \hline
      \BS{0,0,2,2,0}&(0,2,0,0,0,0) &8&4\\ 
    \hline
      \BS{2,2,0,0,0}&(0,0,0,0,4,0)&11&1\\
    \hline
      \BS{4,0,0,0,0}&(1,0,0,2,0,0)&6&1\\
    \hline
      \BS{5,0,0,0,0}&(0,0,0,2,0,0) &6&2\\
    \hline
      \BS{5,0,0,0,0}&(2,0,0,2,0,0) &6&2\\
    \hline
  \end{array}
\end{equation}
The first one of these boundary states corresponds to a $D4$-brane
wrapped around the 
exceptional divisor which is a ruled surface $E$.
\begin{equation}
  r(V)=0,\quad\ch_1=2E,\quad\ch_2(V)=4l-8h\quad\ch_3(V)=-36
\end{equation}
There are only a few results known about semistable sheaves on ruled
surfaces. Their 
application to our problem is work in progress. The second one
corresponds to a $D2$-brane 
wrapped on the 2-cycle $h$ which lies at the intersection of the
divisors $H$ and $L$.
\begin{equation}
  r(V)=0,\quad\ch_1(V)=0,\quad\ch_2(V)=-4h,\quad\ch_3(V)=0
\end{equation}
The fourth one is the $D0$-brane which has already appeared as a stable
sheaf on the $K3$ fiber.
\begin{equation}
  r(V)=0,\quad\ch_1(V)=0,\quad\ch_2(V)=0,\quad\ch_3(V)=-12
\end{equation}
As in the case of some of the one-parameter models it describes two
different $D0$-branes, 
each of them having 3 moduli. Next we are going to consider bound states
of $D0$- and $D6$-branes. 
They are described by the boundary states in the third and and fifth row
of the table. 
The first of these is completely analogous to the one observed
in~\cite{9910172} for 
the elliptic fibration.
\begin{equation}
  r(V)=1,\quad\ch_1(V)=0,\quad\ch_2(V)=0,\quad\ch_3(V)=-12
\end{equation}
The supersymmetric boundary state in the Gepner model decays into a
non-super-symmetric 
configuration of D-branes at the large volume limit. The authors
of~\cite{9910172} have 
given an interesting interpretation from the point of view of the mirror
$\widehat{X}$ 
which we shortly repeat here. In~\cite{9606040} it has been proposed
that mirror 
symmetry is T-duality when $X$ and $\widehat{X}$ admit special
Lagrangian fibrations 
with T-dual $T^3$, $\widehat{T}^3$ fibers. Mirror symmetry now maps the
$D6$-brane 
on $X$ to $D3$-branes wrapping the base $B$ of the fibration while the
$D0$-branes are 
mapped to $D3$-branes wrapping the $\widehat{T}^3$ fiber. The above
decay process now 
tells us that the corresponding homology class $B + \widehat{T}^3$
should not support a 
special Lagrangian cycle in a neighborhood of the large complex
structure limit. 
It should support it instead in a region of the moduli space of
$\widehat{X}$ which 
is mapped to a neighborhood of the Gepner point of $X$ by mirror
symmetry. These are 
the phase transitions of special Lagrangian cycles under the deformation
of the complex 
structure of $\widehat{X}$ which have been studied in~\cite{9907013}
and~\cite{9908135}. 
In our example we find another possibility when we consider the boundary
state in the 
last row.
\begin{equation}
  r(V)=2,\quad\ch_1(V)=0,\quad\ch_2(V)=0,\quad\ch_3(V)=-12
\end{equation}
It is also a bound state of a $D0$- and a $D6$-brane, but now there are
two vacua which 
can be interpreted as two bound states of one $D6$- and one $D0$-brane
with three moduli 
each.

For $\mP_{1,1,2,2,2}^4[8]$ the following boundary states correspond to a
D-brane wrapped 
on the $K3$ fiber. 
\begin{equation}
  \begin{array}{|c|c|c|c|c|}
    \hline
      \textrm{$L$-orbit}&\textrm{Mukai vector
$v=(n_4^2,n_2^1,-n_4^2-n_0)$}&
\textrm{\#moduli}&\textrm{\#vacua}&\textrm{$d$}\\
    \hline
      \BS{1,0,0,0,0}&(3,-4,1)\;(-3,8,-3)\;(1,-4,3)\;(-1,0,-1)&1&1&0 \\
    \hline
      \BS{3,0,0,0,0}&(0,4,-2)\;(2,-4,0)&7&1&6 \\
    \hline
      \BS{3,0,1,0,0}&(2,-8,2)\;(2,0,-2)&14&2&10 \\
    \hline
      \BS{3,0,1,1,0}&(4,-8,0)\;(0,8,-4)&28&4&18 \\
    \hline
      \BS{3,0,1,1,1}&(4,0,-4)\;(4,-16,4)&56&8&34 \\
    \hline
  \end{array}
\end{equation}
Again, all these boundary states satisfy the BPS condition. There are no
single $D0$-branes, 
no $D4$-branes wrapped on the exceptional divisor $E$, nor are there
bound states of $D0$- 
and $D6$-branes.

\acknowledgments{I would like to thank S. Korden, N. Quiroz, and S.
Theisen for 
fruitful discussions, C. R\"omelsberger for clarifying remarks and
helpful comments 
on the manuscript.}

\begin{appendix}

\section{Analysis of the periods of the degree 12 hypersurface in
$\mP_{1,1,2,2,6}^4$}
\label{sec:Candelas}
\setcounter{equation}{0}
Here we perform the calculation of the monodromy matrices for the model
$\mP_{1,1,2,2,6}^4[12]$ 
according to~\cite{9308083}. The fundamental period is given by
\begin{eqnarray}
  \label{eq:right}
  \varpi_0(\psi,\phi)&=&\sum_{r,s=0}^{\infty}\frac{(12r+6s)!(-2\phi)^s}
{(6r+3s)!((2r+s)!)^2(r!)^2s!(12\psi)^{12r+6s}}\\
                     &=&\sum_{n=0}^{\infty}\frac{(6n)!(-1)^n}
{(n!)^3(3n)!(12\psi)^{6n}}u_n(\phi),\qquad\qquad
\left|\frac{\phi\pm1}{864\psi^6}\right| < 1
\end{eqnarray}
where
\begin{eqnarray}
  u_n(\phi) &=&
(2\phi)^n\sum_{r=0}^{[\frac{n}{2}]}\frac{n!}{(r!)^2(n-2r)!(2\phi)^{2r}}
\end{eqnarray}
After extension of the definition of $u_n(\phi)$ to complex values $\nu$
for $n$ we 
write the period as an integral of Barne's type in order to obtain an
expression which is 
valid for small $\psi$.
\begin{eqnarray}
  \varpi_0(\psi,\phi)&=&\frac{1}{2\pi i}\int_C
\diff{}{\nu}\frac{\Gamma(6\nu+1)\Gamma(-\nu)}
{\Gamma(3\nu+1)\Gamma^2(\nu+1)}(12\psi)^{-6\nu}u_{\nu}(\phi),\\&& -\pi < 
\textrm{arg}\left(\frac{864\psi^6}{\phi\pm 1}\right) < \pi
\end{eqnarray} 
For $\left|\phi\pm1\right| < \left|864\psi^6\right|$ we can close the
contour to the right 
and obtain~(\ref{eq:right}) as a sum of the residues of the poles of
$\Gamma(-\nu)$. 
For $\left|\phi\pm1\right| > \left|864\psi^6\right|$ the contour can be
closed to the 
left giving a sum over the residues of the poles of $\Gamma(6\nu+1)$.
Defining
\begin{equation}
  \varpi_j(\psi,\phi) = \cA^j\varpi_0(\psi,\phi) =
\varpi_0(\alpha^j\psi,(-1)^j\phi)
\end{equation}
with $\alpha^{12} = 1$ we find
\begin{equation}
  \varpi_j(\psi,\phi) = -\frac{1}{6}\sum_{m=1}^{\infty}
\frac{(-1)^m\alpha^{jm}\Gamma(\frac{m}{6})}
{\Gamma(m)\Gamma(1-\frac{m}{2})\Gamma^2(1-\frac{m}{6})}(12\psi)^m
u_{-\frac{m}{6}}((-1)^j\phi)
\end{equation}
The factors $\Gamma(1-\frac{m}{2})$ and $\Gamma(1-\frac{m}{6})$ in the
denominator determine the 
relations between the periods as follows
\begin{equation}
  \varpi_j+\varpi_{j+6}=0
\end{equation}
Hence we have for $\cA:\varpi \longrightarrow A_G\varpi$
\begin{equation}
  \label{eq:AG}
  A_G = \left(\begin{array}{cccccc}
0&\hspace{1.7mm}1\hspace{1.7mm}&\hspace{1.7mm}0\hspace{1.7mm}&\hspace{1.7mm}0\hspace{1.7mm}&\hspace{1.7mm}0\hspace{1.7mm}&\hspace{1.7mm}0\hspace{1.7mm}\\\noalign{\medskip}0&0&1&0&0&0\\\noalign{\medskip}0&0&0&1&0&0
\\\noalign{\medskip}0&0&0&0&1&0\\\noalign{\medskip}0&0&0&0&0&1
\\\noalign{\medskip}-1&0&0&0&0&0\end{array}\right)
\end{equation}
To compute the monodromy of these periods around $\phi=1$ we have to
continue the $\varpi_j$ 
to large values of $\psi$. Following~\cite{9308083} we write the
variable of summation as $m=6n+r$
\begin{eqnarray}
  \varpi_{2j} &=&
-\frac{1}{6\pi^3}\sum_{r=1}^5(-1)^r\alpha^{2jr}\sin\left(\frac{\pi
r}{2}\right)
\sin^2\left(\frac{\pi r}{6}\right)\xi_r\\
  \varpi_{2j+1} &=&
-\frac{1}{6\pi^3}\sum_{r=1}^5(-1)^r\alpha^{(2j+1)r}\sin\left(\frac{\pi
r}{2}
\right)\sin^2\left(\frac{\pi r}{6}\right)\eta_r
\end{eqnarray}
where 
\begin{eqnarray}
\xi_r(\psi,\phi)&=&\frac{1}{2i}\int_C\diff{}{\nu}\frac{\Gamma^3(-\nu)\Gamma(-3\nu)}{\Gamma(-6\nu)}
(12\psi)^{-6\nu}\frac{u_\nu(\phi)}{\sin\left(\pi(\nu+\frac{r}{6})\right)}\\
\eta_r(\psi,\phi)&=&-\frac{1}{2i}\int_C\diff{}{\nu}\frac{\Gamma^3(-\nu)\Gamma(-3\nu)}{\Gamma(-6\nu)}
(12\psi)^{-6\nu}\frac{u_\nu(\phi)\sin\left(\pi(\nu+\frac{r}{6})\right)+u_\nu(-\phi)\sin(\frac{\pi
r}{6})}
{\sin\left(\pi(\nu+\frac{r}{6})\right)\sin(\pi \nu)}
\end{eqnarray}
An analogous computation to that in~\cite{9308083} gives for $\cB:
\varpi \longrightarrow B_G\varpi$
\begin{equation}
  B_G=\left(\begin{array}{cccccc}
1&\hspace{1.7mm}0\hspace{1.7mm}&0&\hspace{1.7mm}0\hspace{1.7mm}&0&0\\\noalign{\medskip}1&0&0&0&-1&1
\\\noalign{\medskip}-1&1&1&0&1&-1\\\noalign{\medskip}0&0&1&0&2&-2\\\noalign{\medskip}0&0&-1&1&-1&2
\\\noalign{\medskip}0&0&0&0&0&1
\end{array}\right)
\end{equation}
To compute the monodromy around the conifold point we repeat again the
steps of~\cite{9308083} and 
obtain for $\cT:\varpi \longrightarrow T_G\varpi$
\begin{equation}
  T_G = \left( 
  \begin{array}{cccccc}
2&-1&\hspace{1.7mm}0\hspace{1.7mm}&\hspace{1.7mm}0\hspace{1.7mm}&\hspace{1.7mm}0\hspace{1.7mm}&\hspace{1.7mm}0\hspace{1.7mm}\\\noalign{\medskip}1&0&0&0&0&0\\\noalign{\medskip}-1&1&1&0&0&0\\\noalign{\medskip}-2&2&0&1&0&0
\\\noalign{\medskip}2&-2&0&0&1&0\\\noalign{\medskip}1&-1&0&0&0&1
  \end{array}
  \right) 
\end{equation}
In~\cite{9308083} the large complex structure limit point in the moduli
space was determined to be 
at the intersection of the two divisors $D_{(1,0)} \leadsto
(A_GT_GB_G)^{-1}$ and $D_{(0,-1)} 
\leadsto (A_GT_G)^{-2}$. If we take $(H,L)$ as the basis for the
K\"ahler cone, i.e. $B+iJ=t_1H+t_2L$, 
then the coordinates $t_i$ can be related to the periods $\varpi_j$ by
\begin{eqnarray}
  t_1 &=& -\frac{1}{2} + \frac{\varpi_2 + \varpi_4}{2\varpi_0}\\
  t_2 &=& \frac{1}{2} + \frac{\varpi_1 - \varpi_2 + \varpi_3 - \varpi_4
+ \varpi_5}{2\varpi_0}
\end{eqnarray}
With the ansatz for the prepotential $\cF$
\begin{equation}
  \cF=-\frac{1}{6}\left(4{t_1}^3 +
6{t_1}^2t_2\right)+\frac{1}{2}\left(\alpha {t_1}^2 + 2\beta t_1t_2 + 
\gamma{t_2}^2\right)+(\delta-\frac{2}{3})t_1 +\varepsilon t_2 +
\textrm{const}
\end{equation}
the constants $\delta$ and $\epsilon$ can be fixed the same way as
in~\cite{9308083}
\footnote{In~\cite{9308083} $\delta$ and $\varepsilon$ appear to have
the wrong sign} 
while the constants $\alpha, \beta$ and $\gamma$ can be chosen
appropriately by an $Sp(6,\mZ)$ 
transformation.
\begin{equation}
  \alpha =
0,\quad\beta=0,\quad\gamma=0,\quad\delta=\frac{17}{6},\quad\varepsilon=1
\end{equation}
This choice differs from the one in~\cite{9308083} and is dictated by,
as has been argued 
in~\cite{9906200}, the fact that the state which becomes massless at the
mirror of the 
conifold point should correspond to the ``pure'' six-brane with large
volume 
charges $Q=(1,0,0,0,0,0)$, following~\cite{9510227} and~\cite{9612181}.
Finally we 
give the expression for the matrix $m$
\begin{equation}
  m=\left(\begin{array}{cccccc}
-1&\hspace{1.7mm}1\hspace{1.7mm}&0&\hspace{1.7mm}0\hspace{1.7mm}&0&0\\\noalign{\medskip}\frac{3}{2}&\frac{3}{2}&\frac{1}{2}&\frac{1}{2}&-\frac{1}{2}
&-\frac{1}{2}\\\noalign{\medskip}1&0&1&0&0&0\\\noalign{\medskip}1&0&0&0&0&0
\\\noalign{\medskip}-\textstyle\frac{1}{2}&0&\textstyle\frac{1}{2}&0&
\textstyle\frac{1}{2}&0\\\noalign{\medskip}\textstyle\frac{1}{2}&\textstyle\frac{1}{2}&
-\textstyle\frac{1}{2}&\textstyle\frac{1}{2}&-\textstyle\frac{1}{2}&\textstyle\frac{1}{2}
  \end {array}\right)
\end{equation}
\newpage
\section{The monodromy matrices around the Gepner point}
\label{sec:AL}
\setcounter{equation}{0}
\begin{equation}
  \begin{array}{cc}
    A_L=\left(\begin{array}{cccc}
-3&-1&-6&4\\\noalign{\medskip}-3&1&3&3\\\noalign{\medskip}1&0&1&-1\\\noalign{\medskip}-1&0&0&1
  \end{array}\right)
    &
    \textrm{ for }\mP_{1,1,1,1,2}^4[6]
    \\
    & \\
    A_L=\left(\begin{array}{cccc}
-3&-1&-4&4\\\noalign{\medskip}-2&1&2&2\\\noalign{\medskip}1&0&1&-1\\\noalign{\medskip}-1&0&0&1
  \end{array}\right)
    &
    \textrm{ for }\mP_{1,1,1,1,4}^4[8]
    \\
    & \\
    A_L=\left(\begin{array}{cccc}
-2&-1&-1&3\\\noalign{\medskip}0&1&1&0\\\noalign{\medskip}1&0&1&-1\\\noalign{\medskip}-1&0&0&1
  \end{array}\right)
    &
    \textrm{ for }\mP_{1,1,1,2,5}^4[10]
    \\
    & \\
    A_L = \left(\begin{array}{cccccc}
-1&0&\hspace{1.7mm}1\hspace{1.7mm}&-2&0&0\\\noalign{\medskip}4&-1&0&
4&-4&-4\\\noalign{\medskip}-2&1&1&-2&4&2\\\noalign{\medskip}1&0&0&1&0&0
\\\noalign{\medskip}-1&0&0&-1&1&1\\\noalign{\medskip}1&0&0&1&0&-1
\end{array}\right)
    &
    \textrm{ for }\mP_{1,1,2,2,2}^4[8]
    \\
    & \\
    A_L = \left(\begin{array}{cccccc}
-1&0&\hspace{1.7mm}1\hspace{1.7mm}&-2&0&0\\\noalign{\medskip}2&-1&0&
2&-2&-2\\\noalign{\medskip}-1&1&1&-1&2&1\\\noalign{\medskip}1&0&0&1&0&0
\\\noalign{\medskip}-1&0&0&-1&1&1\\\noalign{\medskip}1&0&0&1&0&-1
\end{array}\right)
    &
    \textrm{ for }\mP_{1,1,2,2,6}^4[12]
  \end{array}
\end{equation}
These can be obtained from the corresponding matrices in~\cite{9205041}
and~\cite{9308083} by 
applying the linear transformation $M$ in~(\ref{eq:M1}), (\ref{eq:M2}),
(\ref{eq:M3}), 
(\ref{eq:M4}) and~(\ref{eq:M5}) respectively to them.

\end{appendix}

% BIBLIOGRAPHY

\end{document}